\documentclass[12pt,preprint]{aastex}
\usepackage{float, amsmath,verbatim}

\shorttitle{Magnetic field generation via the kinetic Kelvin-Helmholtz instability in unmagnetized scenarios} 
\shortauthors{Alves et al.}

\begin{document}

\title{Large-scale magnetic field generation via the kinetic Kelvin-Helmholtz instability in unmagnetized scenarios} 

\author{E. P. Alves\altaffilmark{1}, T. Grismayer\altaffilmark{1}}
\author{S. F. Martins\altaffilmark{1}, F. Fi\'uza\altaffilmark{1}, R. A. Fonseca\altaffilmark{1,2}, L. O. Silva\altaffilmark{1}}
\altaffiltext{1} {GoLP/Instituto de Plasmas e Fus\~ao Nuclear - Laborat\'orio Associado, Instituto Superior T\'ecnico, Lisbon, Portugal}
\altaffiltext{2} {Instituto Universit\'ario de Lisboa (ISCTE-IUL), Lisbon, Portugal}
\subjectheadings{magnetic field generation -- plasma instabilities -- Kelvin-Helmholtz instability}

\begin{abstract}

Collisionless plasma instabilities are fundamental in magnetic field generation in astrophysical scenarios, but their role has been addressed in scenarios where velocity shear is absent. In this work we show that velocity shears must be considered when studying realistic astrophysical scenarios, since these trigger the collisionless Kelvin-Helmholtz instability (KHI). We present the first self-consistent three-dimensional (3D) particle-in-cell (PIC) simulations of the KHI in conditions relevant for unmagnetized relativistic outflows with velocity shear, such as active galactic nuclei (AGN) and gamma-ray bursts (GRBs). We show the generation of a strong large-scale DC magnetic field, which extends over the entire shear-surface, reaching thicknesses of a few tens of electron skin depths, and persisting on time-scales much longer than the electron time scale. This DC magnetic field is not captured by MHD models since it arises from intrinsically kinetic effects. Our results indicate that the KHI can generate intense magnetic fields yielding equipartition values up to  $\epsilon_B/\epsilon_p \simeq 10^{-3}-10^{-2}$ in the electron time-scale. The KHI-induced magnetic fields have a characteristic structure that will lead to a distinct radiation signature, and can seed the turbulent dynamo amplification process. The dynamics of the KHI are relevant for non-thermal radiation modeling and can also have a strong impact on the formation of relativistic shocks in presence of velocity shears.

\end{abstract}

\section{Introduction}

Magnetic field generation in astrophysical scenarios, such as AGN and GRBs, is not fully understood \citep{colgate01}. While such phenomena are of fundamental interest, they are also closely related to open questions such non-thermal radiation emission and cosmic ray acceleration \citep{bhattacharjee00}. Recently, collisionless plasma effects have been proposed as candidate mechanisms for magnetic field generation \citep{gruzinovwaxman99, medvedev99}, mediating the formation of relativistic shocks via the Weibel instability \citep{weibel59,silva03}. The KHI \citep{dangelo65,gruzinov08,macfadyen09} should also be considered since it is capable of generating large-scale magnetic fields in the presence of strong velocity shears, which naturally originate in energetic matter outbursts in AGN and GRBs, and which are also present whenever conditions for the formation of relativistic shocks exist. These large-scale fields may also be further amplified by the magnetic-dynamo effect \citep{gruzinov08,macfadyen09}.

Recent kinetic simulations have focused on magnetic field generation via electromagnetic plasma instabilities in unmagnetized flows without velocity shears.
3D PIC simulations of Weibel turbulence \citep{silva03,fonseca03,frederiksen04,nishikawa05} have demonstrated subequipartition magnetic field generation. The Weibel instability has been shown to be critical in mediating relativistic shocks \citep{spitkovsky08,martins09}, where a Fermi-like particle acceleration process has also been identified. These works have neglected the role of velocity shear in the flow, which are an alternative mechanism to generate sub-equipartition magnetic fields in relativistic outflows \citep{gruzinov08}. Furthermore, a shear flow upstream of a shock can lead to density inhomogeneities via the KHI which may constitute important scattering sites for particle acceleration. In \cite{macfadyen09}, 3D magnetohydrodynamic (MHD) simulations of KH turbulence are discussed, observing magnetic field amplification due to the KHI. In fact, the KHI setup is routinely used to test MHD models \citep{mignone09, beckwith11}, but the connection between the MHD description and the fully kinetic picture is still missing \citep{gruzinov11}. However, the KHI contains intrinsically kinetic features, and so far 3D fully kinetic ab initio simulations have not been reported.

In this Letter, we present the first self-consistent 3D PIC simulations of the KHI for both subrelativistic and relativistic scenarios of shearing unmagnetized electron-proton plasma clouds. We show that the KHI contains important kinetic features which are not captured in previous 3D MHD simulations \citep{keppens99,macfadyen09,mignone09,beckwith11}, namely the transverse KHI dynamics, and the generation of a large-scale DC magnetic field at the shear region. This large-scale field can reach mG levels for typical parameters of the interaction of a relativistic flow with the interstellar medium (ISM). Furthermore, our generalization of the KHI linear theory \citep{gruzinov08} to include arbitrary density jumps between the shearing flows allows us to conclude that the onset of the instability is robust to this asymmetry. The KHI can therefore operate in shears within the ejecta (similar density flows) and also between the ejecta and the surrounding ISM (relative density ratios of $1-10$), and it will likely operate at the same level (or stronger) than the Weibel instability, even for moderate velocity shears. In Section 2, we explore the density jumps effect on the behavior and features of the instability. The 3D PIC simulation results are presented and discussed in detail in Section 3. In Section 4, we discuss the saturation levels of the magnetic field, and conclusions are drawn in Section 5.

\section{Theoretical analysis} \label{sec:theory}

\begin{figure*} [t]
\centering
\includegraphics[width = 1.\columnwidth]{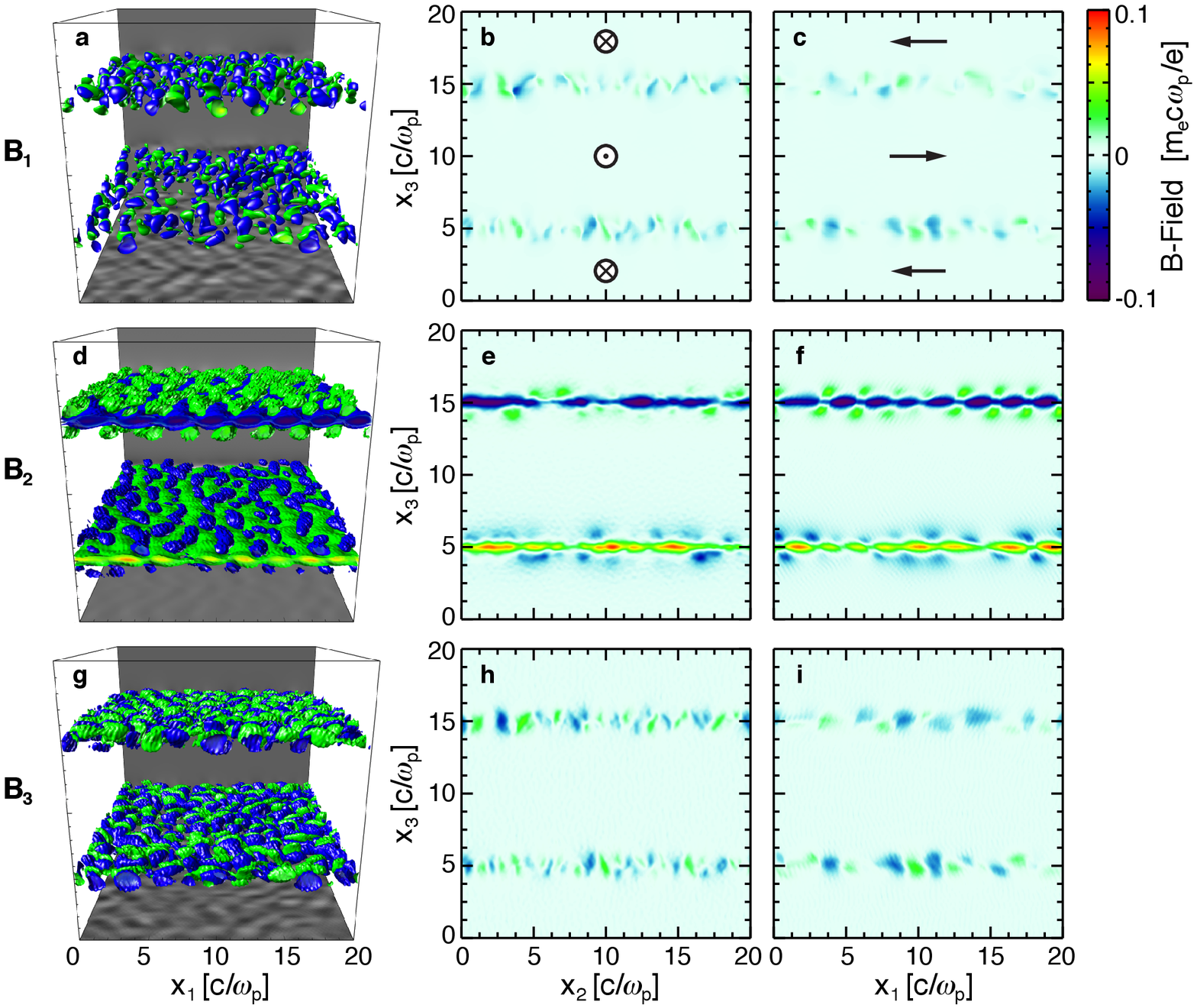}
\caption{Magnetic field structures generated by shearing subrelativistic $e^-p^+$ flows with $\gamma_0 = 1.02$ taken at time $t = 49/\omega_p$. The 3D visualizations (a), (d) and (g) correspond to the magnetic field components $B_1$, $B_2$ and $B_3$, respectively. The 2D slices of the magnetic field intensity (b), (e) and (h) are taken at the centre of the box $x_1 = 10 ~ c/\omega_p$, and slices (c), (f) and (i) are taken at $x_2 = 10 ~ c/\omega_p$.}
\label{fig:3dclas}
\end{figure*}

The KHI linear theory, outlined in \citep{gruzinov08}, is based on the relativistic fluid formalism of plasmas coupled with Maxwell's equations, and was analyzed for the particular case where the two shearing flows have equal densities. Realistic shears, however, are more likely to occur between different density flows \citep{hardee92,krause03}.

We have extended the analysis presented in \citep{gruzinov08} for shearing electron-proton plasma flows with uniform densities $n_+$ and $n_-$, and counter-velocities $+\vec v_0$ and $-\vec v_0$, respectively. Here, the protons are considered free-streaming whereas the electron fluid quantities and fields are linearly perturbed. The dispersion relation for electromagnetic waves is thus:
\begin{eqnarray}
\label{eq1}
		\sqrt{\frac{n_-}{n_+}+\frac{k'^2}{\beta_0^2}-\omega'^2} 
			\left[
				\left (\omega'+k' \right)^2 - \left(\omega'^2-k'^2\right)^2
			\right] + \nonumber \\ 
		\sqrt{1+\frac{k'^2}{\beta_0^2}-\omega'^2}  
			\left[
				\frac{n_-}{n_+} \left (\omega'-k' \right)^2 - \left(\omega'^2-k'^2\right)^2
			\right] = 0,
\end{eqnarray}
where $\omega ' \equiv \omega/\omega_{p+}$ is the wave frequency normalized to the plasma frequency, with density $n_+$, $\omega_{p+} = (4\pi n_+e^2/\gamma_0^3 m_e) ^{1/2}$. $k' \equiv k ~ c / \omega_{p+}$ is the normalized wave number parallel to the flow direction, $\beta_0 = v_0/c$ and $\gamma_0 = ( 1-\beta_0^2) ^{-1/2}$ is the Lorentz relativistic factor; $e$ and $m_e$ are the electron charge and mass, respectively, and $c$ is the speed of light in vacuum. The unstable modes are surface waves, since their transverse wave number is evanescent.

We now consider two scenarios. The similar-density ($n_+/n_- \simeq 1$) scenario, relevant for instance in internal shocks within the relativistic ejecta, where the densities of the shearing regions are comparable, and the $n_+/n_- > 1$ scenario, relevant for external shocks associated with the interaction of the relativistic ejecta with the ISM. In the first ($n_+/n_-=1$), the modes with $| k' |<1$ are unstable, with a maximum growth rate $\Gamma_\mathrm{max}'=\Gamma_\mathrm{max}/\omega_{p+}=1/\sqrt{8}$ for the fastest growing mode $k'_\mathrm{max}=k_\mathrm{max} ~ c / \omega_{p+} = \sqrt{3/8}/\beta_0$. The real part of $\omega'$ vanishes over the range of unstable modes meaning that the unstable modes are purely growing waves. In the density-contrast ($n_+/n_- > 1$) scenario, the shape of the spectrum is conserved for different density ratios, indicating that the general characteristics of the instability are unchanged. As $n_+/n_-$ increases, the bandwidth of unstable modes decreases, $k'_\mathrm{max}$ drifts to larger scales, and the growth rate $\Gamma'_\mathrm{max}$ decreases. Furthermore, the frequency $\omega'$ acquires a real part over the range of the unstable modes, meaning that the growing perturbations propagate. The density-contrast unbalances the interaction between flows such that the growing perturbations are more strongly manifested in the less dense flow, and drift in the direction of this less dense flow. In the same-density case, the interaction between flows is balanced and the perturbations are purely growing.

\begin{figure*} [t]
\centering
\includegraphics[width = 1.\columnwidth]{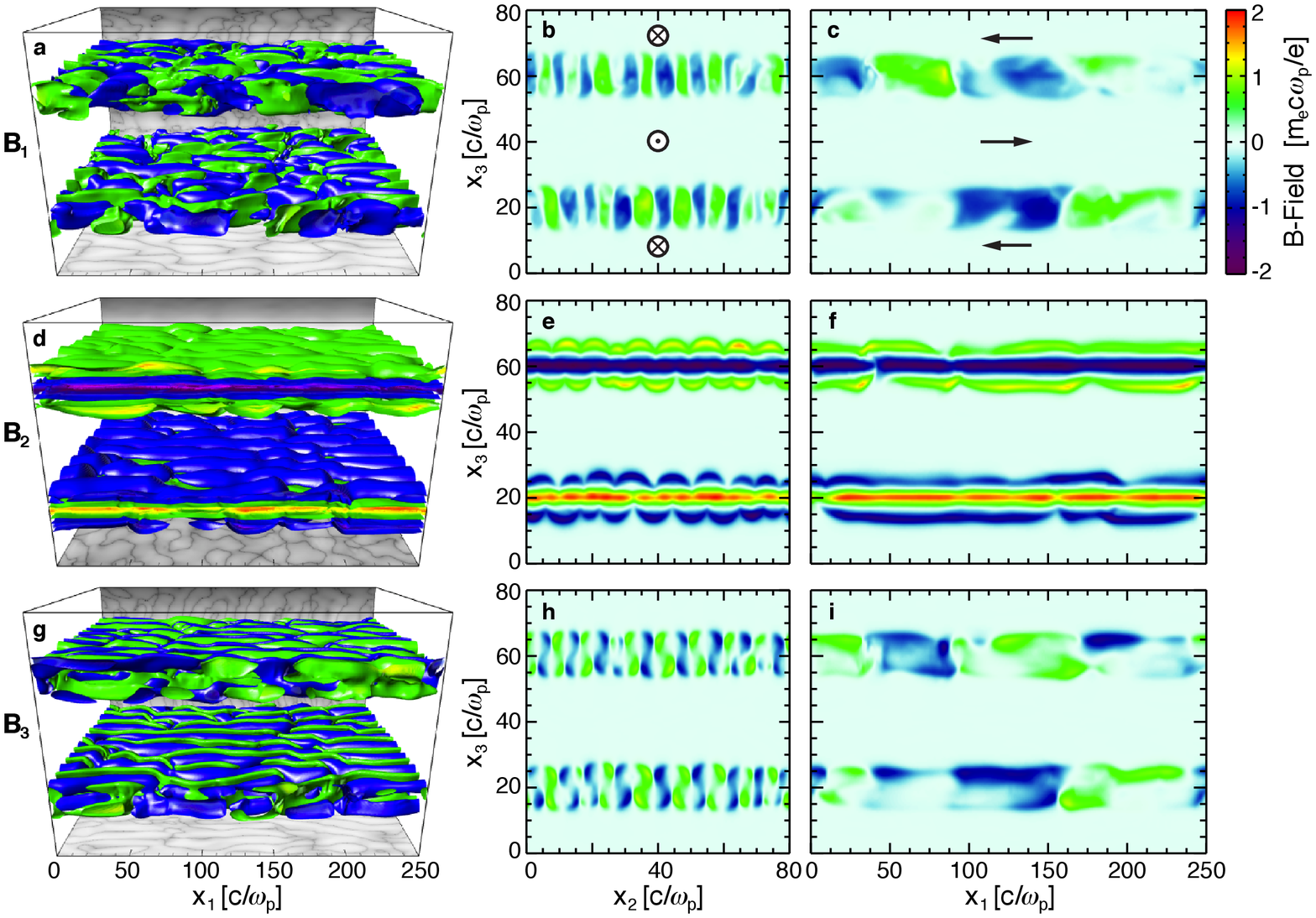}
\caption{Magnetic field structures generated by shearing relativistic $e^-p^+$ flows with $\gamma_0 = 3$ taken at time $t = 69/\omega_p$. The 3D visualizations (a), (d) and (g) correspond to the magnetic field components $B_1$, $B_2$ and $B_3$, respectively. The 2D slices of the magnetic field intensity (b), (e) and (h) are taken at the centre of the box $x_1 = 125 ~ c/\omega_p$, and slices (c), (f) and (i) are taken at $x_2 = 40 ~ c/\omega_p$.}
\label{fig:3drel}
\end{figure*}

The normalized growth rate of the fastest growing mode, $\Gamma'_\mathrm{max}$, decreases with the density contrast between flows. For $n_+/n_- \approx 1$, the growth rate scales approximately as $\Gamma'_\mathrm{max} \propto (n_+/n_-)^{-1/4}$ for both relativistic and non-relativistic shears, whereas in the high density contrasts ($n_+/n_- \gg 1$), the growth rate scales as $\Gamma'_\mathrm{max} \propto (n_+/n_-)^{-1/3}$ for non-relativistic shears (similar to the small cold beam-plasma instability, \citep{oneil71}), and  $\Gamma'_\mathrm{max} \propto (n_+/n_-)^{-1/2}$ for highly relativistic shears. These scalings show that the KHI will compete and operate on the same time-scales as the Weibel instability, and should thus be considered in realistic relativistic outflows where velocity shears are likely to be present.

\begin{figure*} [t]
\centering
\includegraphics[width = 1.\columnwidth]{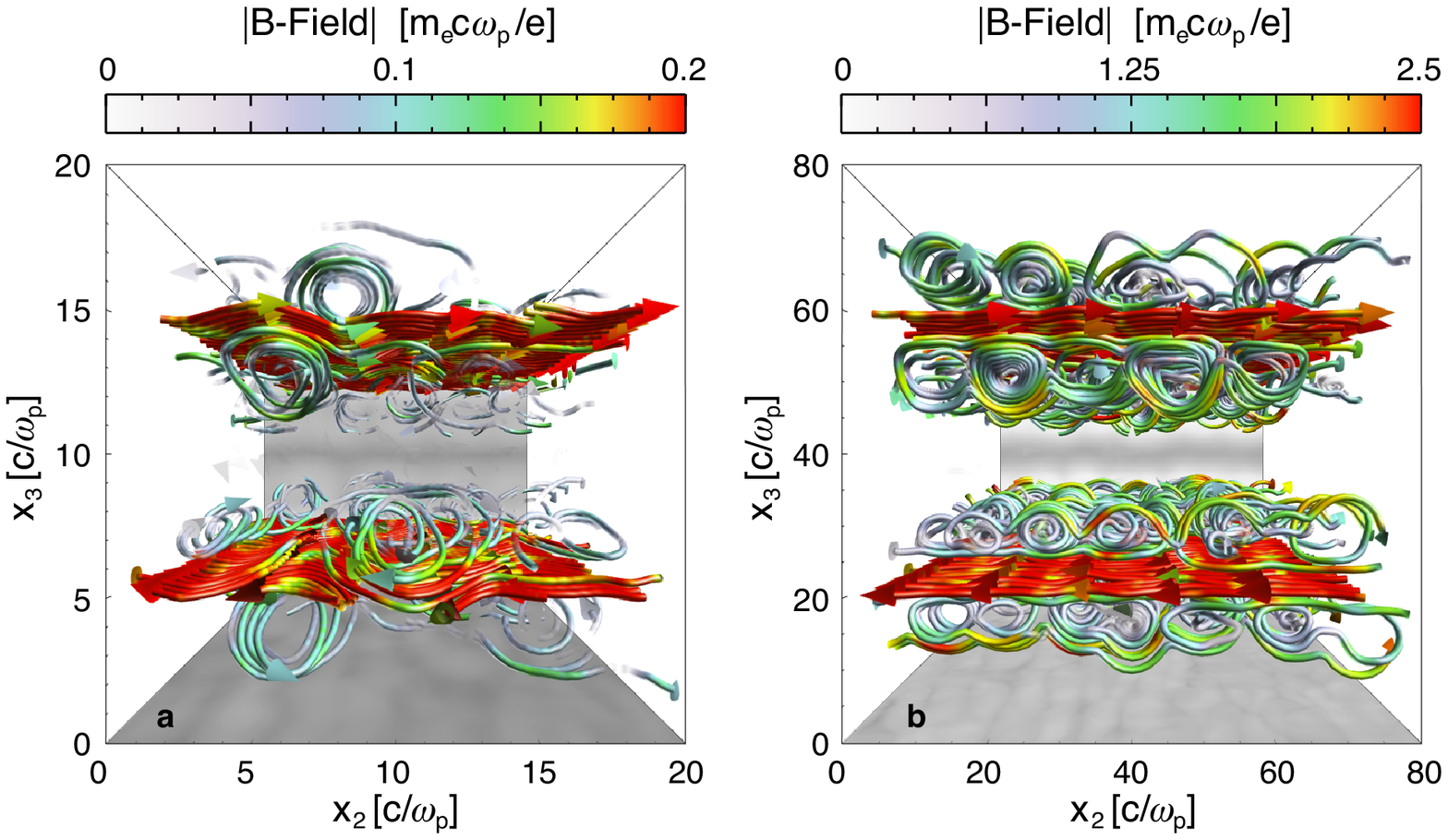}
\caption{Magnetic field lines generated in (a) the subrelativisitc scenario, and (b) the relativistic scenario, at time $t=100/\omega_p$.}
\label{fig:3dlines}
\end{figure*}

\section{3D PIC simulations} \label{sec:sim}

Numerical simulations were performed with OSIRIS, a fully relativistic, electromagnetic, and massively parallel PIC code \citep{fonseca03,fonseca08}, which has been widely used in other astrophysical problems \citep{silva03,medvedev05,medvedev06,martins09}.

We simulate 3D systems of shearing slabs of cold ($v_{0} \gg v_{th}$, where $v_{th}$ is the thermal velocity) unmagnetized electron-proton plasmas with a realistic mass ratio $m_p/m_e=1836$ ($m_p$ is the proton mass), and evolve it until the electromagnetic energy saturates on the electron time-scale. In this Letter, we explore a subrelativistic ($\gamma_0 = 1.02$) and a relativistic ($\gamma_0 = 3$) KHI scenario, in the regime of two plasma slabs of equal density, which is directly relevant to internal shocks where $1 < \gamma_0 < 10$ with comparable density shells \citep{Piran05}. 

The numerical simulations are prepared as follows. The shear flow initial condition is set by a velocity field with $v_0$ pointing in the positive $x_1$ direction, in the upper and lower quarters of the simulation box, and a symmetric velocity field with $-v_0$ pointing in the negative $x_1$ direction, in the middle-half of the box. Initially, the systems are charge and current neutral. In the subrelativistic case, the simulation box dimensions are $20 \times 20 \times 20 ~ ( c/\omega_p )^3$, where $\omega_p = (4\pi n e^2/m_e)^{1/2}$ is the plasma frequency, and we use $20$ cells per electron skin depth ($c/\omega_p$). The simulation box dimensions for the relativistic scenario are  $250 \times 80 \times 80 ~ ( c/\omega_p )^3$, with a resolution of $4$ cells per $c/\omega_p$. Periodic boundary conditions are imposed in every direction. 

The magnetic field structures generated in the subrelativistic and relativistic scenarios are displayed in Figures \ref{fig:3dclas} and \ref{fig:3drel}, respectively, and the magnetic field line topology is illustrated in Figure \ref{fig:3dlines}. In the linear regime, the onset of the fluid KHI, discussed in Section \ref{sec:theory}, occurs in the $x_1x_3$ plane around the shear surfaces, generating a magnetic field normal to this plane, $B_2$. This growing magnetic field component $B_2$ is responsible for rolling-up the electrons to form the signature KH vortices (insets a2 and b2 of Figure \ref{fig:bevol}).
The typical length-scale of the KHI modulations, observed in Figures \ref{fig:3dclas} f and \ref{fig:3drel} f, agrees with the wavelength of the most unstable mode predicted by the fluid theory ($\lambda_\mathrm{max} = 2~c/\omega_p$ and $\lambda_\mathrm{max} \simeq 50~c/\omega_p$ in the subrelativistic and relativistic cases, respectively).
The KHI modulations, however, are less noticeable in the relativistic regime because these are masked by a strong DC component with a magnitude higher than the AC component, which is negligible in the subrelativistic regime.
This DC magnetic field  ($k'=0$), is not unstable according to the fluid model as can be seen in equation (\ref{eq1}).
As the amplitude of the KHI modulations grow, the electrons from one flow cross the shear surfaces and enter the counter-streaming flow. Since the protons are unperturbed, due to their inertia, the current neutrality around the shear surfaces is unbalanced, forming DC current sheets, which point in the direction of the proton velocity.
These DC current sheets induce a DC component in the magnetic field $B_2$.
The DC magnetic field is therefore dominant in the relativistic scenario since a higher DC current is setup by the crossing of electrons with a larger initial flow velocity and also since the growth rate of the AC dynamics is lower by $\gamma_0^{3/2}$ compared with the subrelativistic case. We stress that this DC field is not captured in MHD (e.g. \cite{macfadyen09}) or fluid theories because it results from intrinsically kinetic phenomena. Furthermore, since the DC field is stronger than the AC field, a kinetic treatment is clearly required in order to fully capture the field structure generated in unmagnetized relativistic flows with velocity shear. This characteristic field structure will also lead to a distinct radiation signature \citep{jlmartins10}.

Electron density structures, which have not been reported in MHD simulations to our knowledge \citep{keppens99,macfadyen09,mignone09,beckwith11}, emerge in the plane transverse to the flow direction (insets a1 and b1 of Figure \ref{fig:bevol}),  and extend along the $x_1$ direction forming electron current filaments.
A harmonic perturbation in the $B_3$ component of the magnetic field at the shear surfaces forces the electrons to bunch at the shear planes forming current filaments, which amplify the initial magnetic perturbation $B_3$.
This process is identical to the one underlying the Weibel instability \citep{medvedev99} and leads to the formation of the observed transverse current filaments, along with the exponential amplification of $B_3$ observed in Figure \ref{fig:bevol}.
Figures \ref{fig:3dclas} g and \ref{fig:3drel} g further show that the $B_3$ magnetic field component shares a filamentary structure, underlining its connection in this process.
The electrons undergoing this bunching process, slow down along their initial flow direction. Again, since the protons are unperturbed at these time-scales, DC ($k_{x_2}=0$ mode) current sheets are setup around the shear surfaces in a similar fashion to the longitudinal dynamics previously discussed. These current sheets induce a DC magnetic field in $B_2$ (Figures \ref{fig:3dclas} and \ref{fig:3drel} e), which is responsible for accelerating the evolving filaments across the shear surface, into the counter-propagating flow.
In the relativistic shear scenario, these filaments are strongly rotated due to the high intensity of $B_2$, into the opposing flow, leading to the formation of well defined finger-like density structures, as seen in inset b1 of Figure \ref{fig:bevol}.
These structures are less pronounced in the subrelativistic scenario due to the lower intensity of the DC component of $B_2$ and to the slower Weibel-like electron bunching process (inset a1 of Figure \ref{fig:bevol}).
Meanwhile, the current component $J_3$, associated with the crossing motion of the electron current filaments along the $x_3$ direction, induces the magnetic field component $B_1$ (inset b of Figures \ref{fig:3dclas} and \ref{fig:3drel}).

The growth rate of the magnetic field in the subrelativistic regime agrees with the theoretical $\Gamma_\mathrm{max}$ (Figure \ref{fig:bevol} a), whereas a significant mismatch is found in the relativistic regime (Figure \ref{fig:bevol} b). This deviation is owed to the transverse dynamics of the KHI, which is not taken into account in the 2D theory. We performed 2D simulations matching the longitudinal ($x_1x_3$) and transverse ($x_2x_3$) planes of the 3D setups, in order to assess the independent evolution of the longitudinal and transverse dynamics of the KHI. The results of the 2D simulations of the longitudinal planes were in excellent agreement with the theoretical growth rates for both subrelativistic and relativistic cases, measuring $0.35 ~ \omega_p$ and $0.07 ~ \omega_p$, respectively. In the 2D simulations of the transverse dynamics we measured $\Gamma = 0.1 ~ \omega_p$ and $\Gamma = 0.3 ~ \omega_p$ for the subrelativistic and relativistic scenarios, respectively. Therefore the full 3D evolution of the KHI is mainly determined by the transverse dynamics in the relativistic regime, in contrast to the subrelativistic regime where the longitudinal dynamics is dominant. We stress that these growth rates are faster/comparable to other collisionless plasma processes that would occur in interpenetrating flows \citep{kazimura98,silva03}.

\section{Electron KHI saturation}

\begin{figure*}[!t]
\centering
\includegraphics[width = 1.\columnwidth]{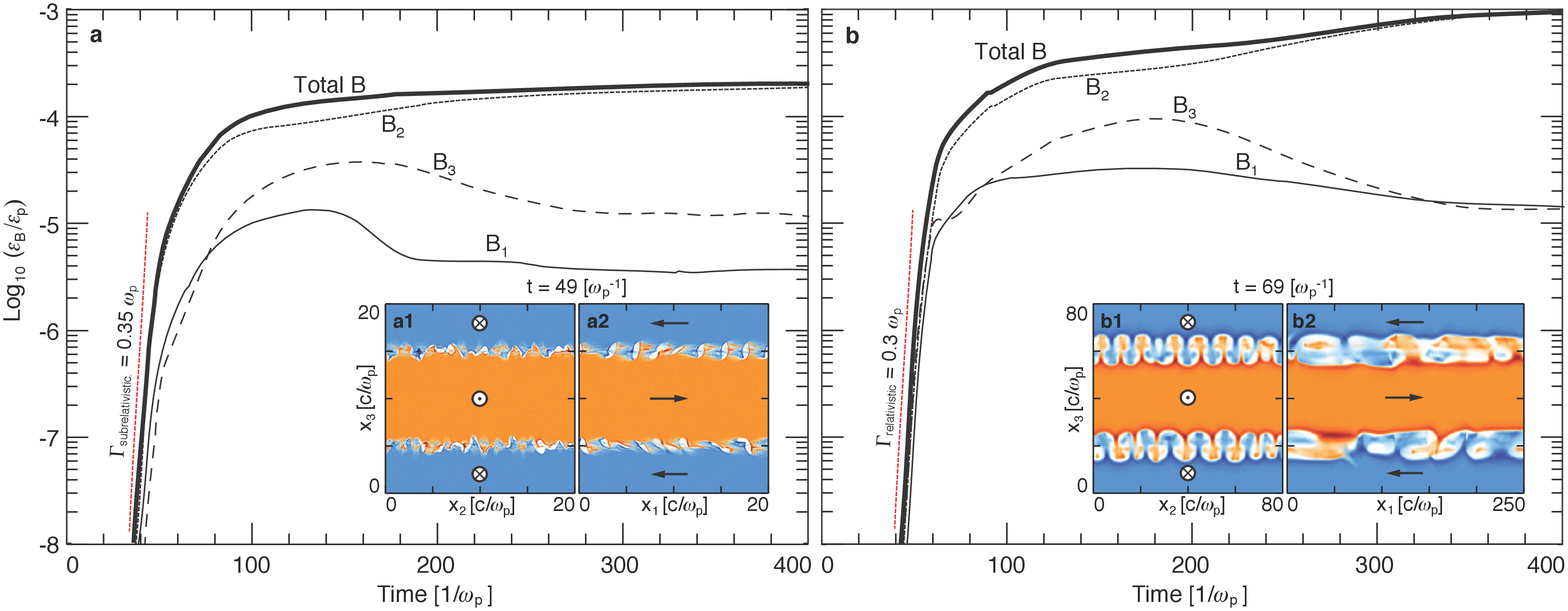}
\caption{Evolution of the equipartition energy $\epsilon_B/\epsilon_p$ for a (a) subrelativistic and (b) relativistic shear scenarios. The contribution of each magnetic field component is also depicted. The insets in each frame represent two-dimensional slices of the electron density at $t=49/\omega_p$ and $t=69/\omega_p$ for the respective case. The red (blue) color represents the electron density of the plasma that flows in the positive (negative) $x_1$ direction. Darker regions in the colormap indicate high electron density, whereas lighter regions indicate low electron density. Slices for insets (a1), (a2), (b1) and (b2) were taken at the center of the simulation box; (a1) and (b1) are transverse to the flow direction, and slices (a2) and (b2) are longitudinal to the flow direction.}
\label{fig:bevol}
\end{figure*}

At later times, the growing KH perturbations begin to interact nonlinearly, ultimately guiding the system into a turbulent state. Eventually, all large-scale shear surfaces in the electron structure become extinct, and the instability saturates.
This stage is reached at roughly $t \simeq 100/\omega_p$ in both subrelativistic and relativistic scenarios (Figure \ref{fig:bevol}).
In this turbulent state, the drift velocity of the electrons vanishes in favor of heating and the magnetic field is mainly sustained by the proton current sheets close to the shear-surfaces.
Here, most of the magnetic field energy is deposited in $B_2$, which has a uniform DC structure that extends throughout the entire shear-surfaces (which can be extremely large-scale in realistic shears), with a characteristic transverse thickness $L_\mathrm{sat} = \alpha ~ c/\omega_p$, where $\alpha$ typically ranges from $5$ to $20$, as measured in the simulations.
The AC modulations in the magnetic field structure are at this time negligible compared to the DC component.
Using Faraday's equation and neglecting the displacement current term, we may estimate the maximum amplitude of the magnetic field as $B_\mathrm{DC}\sim 2\pi e n_0 L_\mathrm{sat} v_0/c$. In the case of relativistic shears, the estimate yields
\begin{equation}
B_\mathrm{DC}\sim 1.6~\alpha~\sqrt{n_0\mathrm{[cm^{-3}]}}~\mathrm{[mG]},
\end{equation}
where $n_0$ is the plasma density. In the case of a relativistic plasma shear of ISM density, $n_0 \simeq 1~ \mathrm{cm^{-3}}$, and assuming $\langle\alpha\rangle \sim 10$, the maximum amplitude of the DC magnetic field is on the order of 10 mG, spreading over the entire shear surface with a thickness of $5\times 10^{9}$ cm. Our simulations confirm this estimate, since they are performed in normalized units (time is normalized to $\omega_p^{-1}$, space to $c/\omega_p$, and magnetic field to $m_e c~\omega_p/e$), showing similar $B_\mathrm{DC}$ levels. The $B_\mathrm{DC}$ estimate can also be used to determine the average equipartition value $\epsilon_B/\epsilon_p$ (ratio of magnetic to initial particle kinetic energy) of the system as
\begin{equation*}
\frac{\epsilon_B}{\epsilon_p}\sim \frac{1}{8}\frac{m_e}{m_p}\frac{\gamma_0+1}{\gamma_0^2} \alpha^2 r,
\end{equation*}
where $\epsilon_p = n_0 (m_p+m_e)c^2(\gamma_0-1) L_{x_1} L_{x_2} L_{x_3}$ ($L_{x_i}$ being the simulation box dimension in the $x_i$ direction), and $r = L_\mathrm{sat}/L_{x_3}$.
We measure from the simulation at saturation $\alpha \simeq 4$ ($r \simeq 0.2$) for the subrelativistic case, yielding $\epsilon_B/\epsilon_p = 4 \times 10^{-4}$, and $\alpha \simeq 15$ ($r \simeq 0.19$) for the relativistic case, yielding $\epsilon_B/\epsilon_p = 1.2 \times 10^{-3}$, which are comparable with the simulations results (Figure \ref{fig:bevol}).
We may also estimate the maximum value of equipartition which is found at the shear surfaces by averaging around the interaction region, which has a typical thickness on the order of $L_\mathrm{sat}$:
\begin{equation*}
\left(\frac{\epsilon_B}{\epsilon_p}\right)_{max}\sim \frac{1}{8}\frac{m_e}{m_p}\frac{\gamma_0+1}{\gamma_0^2} \alpha^2,
\end{equation*}
yielding $2 \times 10^{-3}$ for the subrelativistic case and $7 \times 10^{-3}$ for the relativistic case.
A higher efficiency of conversion of particle kinetic energy to magnetic fields is observed in the relativistic case since the thickness of the proton currents sheets ($L_\mathrm{sat}$) is much higher than in the subrelativistic case.
Most of the energy in the system, however, is still contained in the ions which remain unperturbed at these time-scales. We expect the system to reach higher levels of equipartition once the protons undergo the proton-scale KHI, which would occur at roughly $t_\mathrm{proton-KHI} \simeq 100 ~ (mi/me)^{1/2} /\omega_p \simeq 4000 /\omega_p$. The long-lived large-scale DC magnetic field can thus be sustained up to the proton time-scale which is roughly long enough for prompt GRB emission and early afterglow \citep{Piran99}.

\section{Conclusions}

In this Letter, we present the first self-consistent 3D PIC simulations of the KHI in unmagnetized electron-proton plasmas and analyze their evolution on the electron time-scale. Our results show that the multidimensional physics of the KHI is extremely rich and that kinetic effects play an important role, in particular, in the transverse dynamics of the KHI (which is dominant over the longitudinal KHI dynamics in relativistic shears), and in the generation of a strong large-scale DC magnetic field.

The transverse dynamics of the KHI consists of a Weibel-like electron bunching process, leading to the formation of electron current filaments which are then accelerated across the shear surface, forming finger-like structures. At the electron saturation time-scale, the magnetic field has evolved to a large-scale DC field structure that extends over the entire shear-surface, reaching thicknesses of a few tens  of electron skin depths, and persisting on time-scales much longer than the electron time-scale. This field structure is not captured by MHD or other fluid models, and will lead to a distinct radiation signature. We measure maximum equipartition values of $\epsilon_B/\epsilon_p \simeq 2 \times 10^{-3}$ for the subrelativistic scenario, and $\epsilon_B/\epsilon_p \simeq 7 \times 10^{-3}$ for the relativistic scenario. These equipartition values, which are typically treated as a free parameter in radiation modeling, match the values inferred from GRB afterglows by \citep{panaitescu02}. Moreover, the onset of the KHI is robust to density asymmetries making it ubiquitous in astrophysical settings. 
The KHI may operate in $n_+/n_- \approx 1$ regimes, relevant in GRB internal shocks, and also in $n_+/n_- \gg 1$ regimes, which are important in external shocks.
Future work will address the impact of the KHI in the formation of relativistic shocks in the presence of velocity shears.

\vspace{0.2cm}

\small 

This work was partially supported by the European Research Council ($\mathrm{ERC-2010-AdG}$ Grant 267841) and FCT (Portugal) grants SFRH/BD/75558/2010, SFRH/BPD/75462/2010, and PTDC/FIS/111720/2009. We would like to acknowledge the assistance of high performance computing resources (Tier-0) provided by PRACE on Jugene based in Germany. Simulations were performed at the IST cluster (Lisbon, Portugal), and the Jugene supercomputer (Germany).



\end{document}